\documentclass[graybox]{svmult}
\usepackage{mathptmx,helvet,courier,type1cm}
\usepackage{makeidx,graphicx,multicol}
\usepackage[bottom]{footmisc}

\usepackage{amsmath} % for \quad
\usepackage{amssymb} % for \square

\usepackage{hyperref}
\hypersetup{
	pdftitle={Generalized Uncertainty Principle and Extra Dimensions},
	pdfsubject={Proceedings of the 2016 Karl Schwarzschild Meeting in FIAS, Frankfurt},
	pdfauthor={Sven Koeppel},
	bookmarks=true,
	hidelinks = true
}
\makeindex

\begin{document}

\title*{Generalized uncertainty principle and extra dimensions}
\author{Sven K\"oppel, Marco Knipfer, Maximilano Isi, Jonas Mureika, Piero Nicolini}
\institute{Sven K\"oppel \at Frankfurt Institute of Advanced Studies (FIAS) and Institut f\"{u}r Theoretische Physik, Goethe Universit\"{a}t, Frankfurt am Main, Germany \email{koeppel@fias.uni-frankfurt.de}
\and Marco Knipfer \at Frankfurt Institute of Advanced Studies (FIAS) and Institut f\"{u}r Theoretische Physik, Goethe Universit\"{a}t, Frankfurt am Main, Germany \email{knipfer@fias.uni-frankfurt.de}
\and Maximiliano Isi \at LIGO Laboratory,   California   Institute   of   Technology,   Pasadena,   CA 91125, USA \email{misi@caltech.edu}
\and Jonas Mureika \at Department of Physics, Loyola Marymount University, Los Angeles, CA 90045, USA \email{jmureika@lmu.edu}
\and Piero Nicolini \at Frankfurt Institute of Advanced Studies (FIAS) and Institut f\"{u}r Theoretische Physik, Goethe Universit\"{a}t, Frankfurt am Main, Germany
\email{nicolini@fias.uni-frankfurt.de}
}

\maketitle

\vspace{-1cm} % hacky solution to save space
\abstract{The generalized uncertainty principle (GUP) is a modification of standard quantum mechanics due to Planck scale effects. The GUP has  recently been used to improve the short distance behaviour of classical black hole spacetimes by invoking nonlocal modifications of the gravity action.  We present the problem of extending such a GUP scenario to higher dimensional spacetimes and we critically review the existing literature on the topic.}

\section{Generalized uncertainty principle and black holes}

\begin{figure}[t]
\begin{center}

%\sidecaption[t]
%\includegraphics[scale=.55]{completeness-real-gup.pdf}
%
\includegraphics[width=6.5cm]{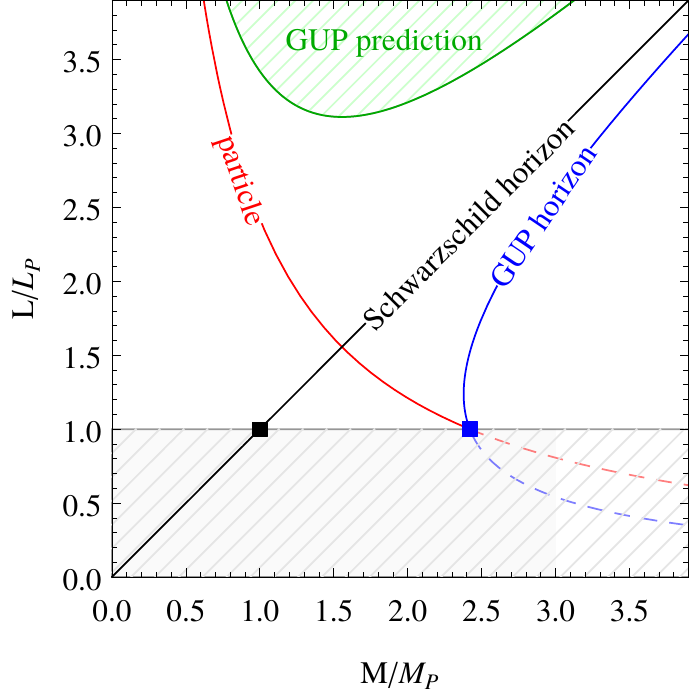}

\caption{Length vs. mass in the Planck region: 
The black line shows the Schwarzschild horizon length scale $L\sim M$, the red line the Compton wavelength $L \sim 1/M$ in Planck units. 
%This figure is well-known \cite{aurilia,dvali2010,carr14}. 
%The GUP (eq. \ref{eq:hotremnants}) predicts a minimal length (green point). 
The GUP green curve interpolates the two curves and predicts a minimal length but does not resolve the phase ambiguity (black spot). 
The GUP-inspired black holes \cite{IMN13} modify the intersection of the Compton/horizon curves due the introduction of a cold remnant (blue spot). In the resulting diagram, the two phases (particles and black holes) are unambiguously separated.  %and smoothly connected by a cross-over at the Planck scale.
}\label{plotgup}
\end{center}

\end{figure}

Gravitation plays no conventional role in quantum mechanical systems.   Atoms are dominated by electromagnetic forces,  with nuclear forces becoming relevant at the smaller sub-atomic scales. It is, however, interesting to ask how quantum mechanics deviates from its standard formulation if the above systems were subject to gravitational interactions.  A full answer to this question would require a quantum theory of gravity, whose formulation is probably one of the biggest problems in fundamental physics. There is nevertheless a ``side effect'' of quantum gravity that one can estimate in a semiclassical way.  For instance, one can consider a non-vanishing Newtonian gravitational interaction between the photon and the electron in Heisenberg's microscope \textit{Gedankenexperiment} \cite{Adl10}. As a result one finds a modification of standard commutation relations \cite{Ven86,ACV89,ACV93,Mag93,KMM95}% \cite{veneziano86,amati95,maggiore95,kmm96}
 \begin{equation}
 [x^i, p_j] = i\, \hbar\, \delta^i_{\, j} \, (1+f( \vec{p}^2)),
 \label{eq:commrel}
 \end{equation}
where the function $f$ is customarily assumed as $f(\vec{ p}^2)\simeq \beta \vec{ p}^2+\dots$ to first order.  Interestingly, the parameter $\beta$ turns out to be a natural ultraviolet cutoff, since the corresponding uncertainty relations prevent better spatial resolution than $\sqrt{\beta}$, %encapsulated by the relation \eqref{eq:commrel}
\begin{equation}
\Delta x \Delta p \geq \frac \hbar2 (1+\beta(\Delta p)^2).
\end{equation}
The above relation is the Generalized Uncertainty Principle (GUP), represented in Fig. \ref{plotgup}. %\cite{veneziano86,amati95,maggiore95,kmm96}.  %Another important consequence of the GUP is the so-called self-completeness. 

%In quantum mechanics, the canonical commutation relation $[x^i, p_j] = i \hbar \delta^i_j $ results in Heisenberg's uncertainty principle $\Delta x \Delta p \geq \frac \hbar2$ between position $x$ and momentum $p$. Adding momentum dependent terms $f(p)$ to get $\Delta x \Delta p \geq \frac \hbar2 (1+f(\Delta p))$ is well known in literature \cite{veneziano86,amati95,maggiore95,kmm96} and describes a theory with a generalized uncertainty principle (GUP). The simplest addition $f(p) = \beta p^2$ with an energy scale $\sqrt{\beta}$ already allows sketching UV-complete gravity as in figure \ref{plotgup}.

The GUP has been studied in a variety of physical systems (for reviews see \cite{SNB12,Hos13,TaM15}), and applied most notably to black holes and their evaporation \cite{Mag93,ChA03,APS01,COY14}. By assuming the emitted particles have momenta uncertainty proportional to the black hole temperature, $\Delta p\sim T$, and position uncertainty proportional to the black hole size, $\Delta x\sim GM$, one ends up with a non-divergent increase of the black hole temperature and vanishing heat capacity at the Planck scale.  Such a scenario for the final stage of the evaporation would suggest the formation of a black hole remnant -- a Planckian size, neutral object that might be considered as a  dark matter candidate \cite{AdS99,APS01}. Unfortunately this particular GUP temperature profile cannot be associated to a surface gravity of any known black hole metric. In addition such remnants would turn to be hot, since their temperature is of the order of the Planck temperature $T_\mathrm{P}\sim 10^{32}$ K.

%\runinhead{Hot Remnants} are an immediate consequence of the reasoning $\Delta x \Delta p \sim 1 + (\Lp \Delta p)^2$ with the solution
%\begin{equation}\label{eq:hotremnants}
%\Delta p_{1,2} \sim \frac{\Delta x}{2 L_P^2} \left(
%1 \pm \sqrt{ 1 - \frac{L_P^2}{\Delta x^2} } \right).
%\end{equation}
%Applied to quantum black hole uncertainties, one can postulate a Hawking temperature $T \sim M^{-1} \sim \Delta p_{1,2}^{-1}$ as was first done by Adler \cite{adler92,adler01}. A hot evaporation endpoint poses questions about stability and coherence of the theory. Instead, we proceed to a more consistent model of deriving a quantum black hole solution taking GUP effects into account.

Against this background, it has recently been noted that GUP effects can be implemented at the level of the spacetime metric by a non-local deformation of the gravitational action \cite{IMN13}. 
%% of the kind
%% \begin{equation}
% S={c^{4} \over 16\pi G}
%%  \int  R{\sqrt {-g}}\,\mathrm {d} ^{4}x\; \rightarrow %S={c^{4} \over 16\pi G}
%%  \int  {\cal F}(R,\, \Box,\, \dots)\, {\sqrt {-g}}\,\mathrm {d} ^{4}x\;,
%%  \label{eq:nlge}
%% \end{equation}
%% where $\Box=\nabla_\mu\nabla^\mu$ is the covariant D'Alembertian and dots stand for higher derivative terms \cite{IMN13}. 
This approach allows for calculating corrections to black hole thermodynamics by a genuine modification of the surface gravity.  In the case of a spherically symmetric, static, and neutral black hole, one can solve the non-local equations and obtain  the metric
\begin{multline}\label{eq:definition-metric}
%multline = multiline equation in package amsmath
\mathrm{d}s^2 = 
-\left(1 - \frac{2G\mathcal{M}(r)}{r}\right)\mathrm{d}t^2
+\left(1 - \frac{2G\mathcal{M}(r)}{ r}\right)^{-1} \mathrm{d}r^2
+ r^2 \mathrm{d}\Omega^2 .
%\\
%\text{ with the modified Newton's potential }
%\Phi(r)=G\mathcal{M}(r) / r.
\end{multline}
Here $\mathcal{M}(r)$ takes into account the spread of matter that is no longer concentrated in a point, as conventionally occurs in the Schwarzschild case \cite{Nic09,BaN93,BaN94,DeB08,SpS17}.
For a specific profile of the nonlocal action, the mass distribution $\mathcal{M}(r)$ reproduces the ultraviolet smearing predicted by the GUP.  The final result then reads
\begin{equation}\label{eq:metric}
\begin{aligned}
\mathcal{M}(r) &=  M \, \gamma(2,\,r/\sqrt{\beta})= M\, \left( 1 - e^{-r/\sqrt{\beta}} - (r/\sqrt{\beta}) e^{-r/\sqrt{\beta}} \right),
\end{aligned}
\end{equation}
with $M$ the ADM mass and $\gamma(s,\, x)$ the lower incomplete gamma function.  A compelling feature of the GUP is that it implies a tail of the mass distribution trespassing the event horizon, in close analogy with the leakage of quantum mechanical effects that has been recently invoked to overcome the black hole information paradox \cite{GBU16}.  The above metric features a two-horizon structure and an extremal configuration.  The latter is a zero temperature state that nicely fits into the cold dark matter paradigm.  Interestingly, the presence of the remnant makes the metric consistent with the expected self-complete character of gravity \cite{DFG11,DGG11,DvG10,DvG12,AuS13,AuS13b,MuN12,Car14,CMN15,
SpA11,NiS12,FKN16}.
 Rather than a complete evaporation, the  black hole asymptotically approaches the remnant configuration, preventing the exposure of length scales smaller of $\sqrt{\beta}$ (see Fig. \ref{plotgup}).

\section{Extra dimensions and the Heisenberg microscope}

Terascale quantum gravity is a formulation proposed to address the weak hierarchy problem of the Standard Model by assuming the existence of additional spatial dimensions \cite{RuS83,RuS83b,AAD98,ADD98,ADD99,BaF99,RaS99a,RaS99b,ACD01,Gog99,Gog00,Gog02}. According to the Arkani-Hamed/Dimopoulos/Dvali (ADD) model, the spacetime is endowed with $N-3$ extra dimensions which are compactified at a length scale $R\sim 1$ mm or smaller. Gravity is the only fundamental interaction able to ``see'' the additional dimensions, which would become relevant only for high energy events (\textit{i.e.}, at the TeV or higher). If this is the case, GUP effects should be expected to set in at these energy scales rather than at the usual Planck scale. It is therefore natural to consider the higher dimensional extension of the non-local action proposed in \cite{IMN13} and derive the related black hole solution. To reach this goal, however, one has to face a potential ambiguity that we illustrate below.

According to the Kempf-Mangano-Mann (KMM) model \cite{KMM95}, the Hilbert space representation of the identity reads
\begin{equation}\label{eq:illkempf}
1 = \int \frac{\mathrm{d}^N p}{1 + \beta {\vec p}^2} \mid p \rangle \langle p \mid ,
\end{equation}
with $\vec p$ an $N$-dimensional spatial vector. This can be interpreted as saying that, while momentum operators preserve their standard character, position operators do not have physical eigenstates, as one expects in the presence of a minimal length $\sqrt{\beta}$. Accordingly the integration measure in momentum space is squeezed in the ultraviolet regime as follows 
\begin{equation}
\mathrm{d}{\cal V}_p \equiv \frac{\mathrm{d}^N p}{1 + \beta {\vec p}^2} \underset{\beta {\vec p}^2\gg 1}{\overset{}{\approx}}
{\vec p}^{N-3}\ \mathrm{d}p.
\label{eq:intvol}
\end{equation}
We note %
%$N$ is kept arbitrary, since 
the GUP correction becomes less and less important with increasing $N$.

The above profile of GUP corrections can be used to improve the higher dimensional Newtonian potential. This can offer a first taste of the repercussions of GUP, even before extending the action proposed in \cite{IMN13} to the higher dimensional case. To reach this goal one has to consider the exchange of virtual massless scalars between two static bodies at distance $r=|\vec{x}|$. The static gravitational potential is the Fourier transform of the massless scalar propagator, 
 \begin{equation}
 \Phi(r)=-\frac{G_{(N)}M}{(2\pi )^{N}}  \int {\mathrm{d}{\cal V}_p}\;\;
  D\left(p\right)\mid _{p_{0}=0}\;\exp \left(i{\vec {p}}\cdot
   {\vec {x}}\right),
\end{equation} 
where the integration measure has been deformed as in \eqref{eq:intvol}. The net result reads
\begin{equation}
 \frac{\Phi(r)}{G_{(N)} M}   
    = \underbrace{\pi^{1-\frac{N}{2}}\Gamma\left( \frac{N}{2}-1 \right)\left(
\frac{1}{r} \right)^{N-2}}_\text{classical potential}
-\underbrace{ \frac{2}{(2\pi \, r\,\sqrt{\beta})^{\frac{N}{2}-1}}\ K_{\frac{N-2}{2}}\left( \frac{r}{\sqrt{\beta}} \right)}_\text{GUP
corrections}
\end{equation}
for $r<R$. The short distance behaviour of the above potential is regular at the origin only for $N=3$, \textit{i.e.},  $\Phi=1/\sqrt{\beta}$ as $r\to 0$. For $N>3$ GUP corrections are suppressed and cannot improve the potential.

There are, however, other proposals. As stated in the introduction, the GUP arises from the inclusion of gravitational effects in quantum mechanics. This is the case of Heisenberg's microscope.  Additional dimensions should not disrupt the reasoning that leads to the GUP \cite{Adl10,Car14}.
As for $N=3$ we identify two terms for the spatial uncertainty $\Delta x \sim \Delta x_\mathrm{C} + \Delta x_\mathrm{g}$, one coming from the Compton wavelength of a particle $\Delta x_\mathrm{C} \sim \lambda \sim 1/\Delta p< R$ and one due to the gravitational potential in the compact higher dimensional space. %Schwarzschild-Tangherlini potential,
Specifically, the photon not only illuminates the electron but also exerts a gravitational force on it. The resulting acceleration causes the electron to be displaced by 
\begin{equation}
\Delta x_\mathrm{g}
\sim G_{(N)} \frac{M_{\rm eff}}{r^{N-1}} \left(\frac{r^2}{c^2}\right)
\sim G_{(N)} \frac{\Delta p}{r^{N-3}}\leadsto \Delta x_\mathrm{g}^{N-2}
\sim L_{(N)}^{N-1} \Delta p,
\end{equation}
where $M_{\rm eff}=h/(\lambda c)$ is the photon effective mass, $G_{(N)}$ is the higher dimensional Newtonian constant, and $L_{(N)}$ is the new fundamental length scale that replaces the Planck length.
In the above derivation we assumed the interaction distance $r\sim \Delta x_\mathrm{g}<R$.
The above Gedankenexperiment motivates a \emph{modified} higher dimensional GUP 
\begin{equation}\label{eq:modGUP}
\Delta x \Delta p \geq \frac{\hbar}{2} \left( 1 + \left( \sqrt{\beta}\, \Delta p \right)^{\frac{N-1}{N-2}} \right),
\end{equation}
where we assumed $\sqrt{\beta}\sim L_{(N)}$.  Eq. \eqref{eq:modGUP} is consistent with what proposed
in \cite{ScC03,Maz13,Car13,Car14,LaC15,LaC16} and cleanly reproduces the higher dimensional Schwarzschild radii for energies above the terascale. On the other hand, such a proposal fails to improve the Newtonian potential and predicts GUP corrections even milder than those of the KMM model for $N>3$.

Alternatively, one can revise the basic reasoning behind Heisenberg's microscope in higher dimensional space. Following Fig. \ref{plotgup}, one can approach the quantum gravity scale from the left. In such a sub-Planckian regime, the gravitational corrections are still sub-leading, \textit{i.e.}, $\Delta x_\mathrm{g}< \Delta x_\mathrm{C}$. Accordingly, one can assume that at the leading order, the typical interaction distance is controlled by the Compton wavelength $r\sim \Delta x_\mathrm{C}\sim\lambda<R$. As a result, one finds a spatial uncertainty
\begin{equation}
\Delta x_\mathrm{g}
\sim G_{(N)} \frac{M_{\rm eff}}{r^{N-1}} \left(\frac{r^2}{c^2}\right)
\sim G_{(N)} \frac{\Delta p}{r^{N-3}}
\sim L_{(N)}^{N-1} \Delta p^{N-2}
\end{equation}
%%% The above Gedankenexperiment motivates a \emph{modified} higher dimensional GUP 
%%% \begin{equation}\label{eq:modGUP2}
%%% \Delta x \Delta p \geq \frac{\hbar}{2} \left( 1 + \left( \sqrt{\beta}\, \Delta p \right)^{N-1} \right),
%%% \end{equation}
%%% where we assumed $\sqrt{\beta}\sim L_{(N)}$. Eq. \eqref{eq:modGUP} is consistent with what proposed
that is consistent with that proposed in \cite{Maz12,DMS15,Maz15}. Interestingly enough, the above relation can also improve the asymptotic behaviour of the momentum integration \eqref{eq:intvol}, as follows %(eq.~\ref{eq:asym}).
\begin{equation}
\mathrm{d}{\cal V}_p \equiv \frac{\mathrm{d}^N p}{1 + (\beta {\vec p}^2)^{\frac{N-1}{2}}} \underset{\beta {\vec p}^2\gg 1}{\overset{}{\approx}}
\ \mathrm{d}p.
\label{eq:intvol3}
\end{equation}
The repercussions of the GUP are no longer dependent on $N$, and the suppression of higher momenta is consistent with the original derivation for $N=3$. The above integration measure \eqref{eq:intvol3} relaxes the condition of reproducing the Schwarzschild radius in the trans-Planckian regime, even if the condition remains valid for distances $r>R$. Such a deviation of the curve from the Schwarzschild radius for $r<R$ is fully legitimate, since the quadratic correction in \eqref{eq:intvol} is known to be the lowest energy correction to standard quantum mechanics and makes sense only for the four dimensional spacetime at scales $r>R$. 

It is natural to expect that at scales $r<R$ where gravity becomes so strong as to probe additional dimensions, the GUP effects also become stronger  as in \eqref{eq:modGUP}. In other words, the picture based on matching a length scale dictated by general relativity \cite{ScC03,Maz13,Car13,Car14,LaC15,LaC16} loses its meaning at an energy regime characterized by string/p-brane effects \cite{AAS02}, noncommutative geometry \cite{KoN10,SSN06} or a variety of non-classical effects \cite{MNT12}. Such a vision is also consistent with a recent proposal aiming to interpret black holes in terms of a pre-geometric, purely quantum mechanical formulation \cite{CGM16,DvG14,SpS17}.

\section{Conclusions}
In this paper, we have reviewed the basic properties of the GUP in three or more dimensions. We have presented a black hole metric derived by a nonlocal action able to reproduce GUP effects \cite{IMN13}. Such a metric overcomes the usual limitations one encounters when considering GUP effects in Hawking radiation \cite{APS01}. Specifically, the new metric allows for the presence of cold remnants and the derivation of the black hole temperature in terms of surface gravity. In the second part of the paper, we provided an analysis of the GUP in higher dimensional spacetime. We showed there is a potential ambiguity in the deformation of the measure in momentum space. We highlighted that current proposals are unable to reproduce a consistent cutoff to improve the bad short distance behaviour of gravity in higher dimensional spacetimes \cite{ScC03,Maz13,Car13,Car14,LaC15,LaC16}. As a possible resolution to such a issue, we revised the reasoning of Heisenberg's microscope in higher dimensional spacetimes. We proposed an improved version of the GUP valid at length scales below the extra-dimensional compactification radius. Our findings are consistent with previous approaches in the literature \cite{Maz12,DMS15,Maz15}. We plan to study the repercussions of such a proposal on black hole physics in a future investigation. 

%%% The Generalized Uncertainty Principle in four dimensions allows the formulation of a black hole solution in the framework of self-complete gravity, featuring black hole remnants. In the scenario of large extra dimensions, we lost this property and the solutions of the same GUP are indistinguishable from the classical Schwarzschild-Tangherlini metric.

%%% Extra dimensions are different: In order to overcome the pathologies, we recognize several ways, i.a.
%%% \begin{itemize}
%%% \item Improve the generalization of the Heisenberg uncertainty principle. Other than that mentioned, several proposals exist, ie. in \cite{maziashvili}
%%% \begin{equation}
%%% \Delta x \Delta p \geq \frac{\hbar}{2} \left( 1 + \left( \sqrt{\beta} p \right)^{(2+n)/(1+n)} \right),
%%% \end{equation}
%%% a definition which is however not compatible with the Gedankenexperiment of Heisenberg's microscope is not sufficient to encounter the singularities.
%%% \item Improve the nonlocal gravity theory: By including higher derivative terms in the nonlocal operator, better properties can be expected.
%%% \item Propose a nonsingular model with the methods of model engineering, as done in \cite{hayward,koeppel}. This method can be applied to reverse-engineer generalizing terms in the uncertainty principle.
%%% \end{itemize}

%%% \bibliographystyle{spphys}	% (uses file "inspires_t_n" "plain.bst", "apsrev.bst", "apsrmp.bst", "unsrt.bst" "JHEP.bst" "unsrtnat.bst" iopart-num, habbrv, hacm, halpha, hapalike, hieeetr )
%%%\bibliography{refs}		% expects file "myrefs.bib"

\begin{acknowledgement}
This work has partially been supported by the project ``Evaporation of the
microscopic black holes'' of the German Research Foundation (DFG) under the grant NI 1282/2-2.
\end{acknowledgement}

% referenc.tex
% References for the KSM-GUP-Proceeding.
%\biblstarthook{My references are the ones as given in the slides.}
% good source: <thesis>/build/thesis.bbl
%

\end{document}